# Generalized Sagnac Effect

Ruyong Wang, Yi Zheng, and Aiping Yao
*St. Cloud State University, St. Cloud, Minnesota 56301, USA*

Experiments were conducted to study light propagation in a light waveguide loop consisting of linearly and circularly moving segments. We found that any segment of the loop contributes to the total phase difference between two counterpropagating light beams in the loop. The contribution is proportional to a product of the moving velocity $\boldsymbol{v}$ and the projection of the segment length $\Delta \boldsymbol{l}$ on the moving direction, $\Delta\phi = 4\pi\boldsymbol{v}\cdot\Delta\boldsymbol{l}/c\lambda$. It is independent of the type of motion and the refractive index of waveguides. The finding includes the Sagnac effect of rotation as a special case and suggests a new fiber optic sensor for measuring linear motion with nanoscale sensitivity.

The Sagnac effect [1] shows that two counterpropagating light beams take different time intervals to travel a closed path on a rotating disk, while the light source and detector are rotating with the disk. When the disk rotates clockwise, the beam propagating clockwise takes a longer time interval than the beam propagating counterclockwise, while both beams travel the same light path in opposite directions. The travel-time difference between them is $\Delta t = 4A\Omega/c^2$, where $A$ is the area enclosed by the path and $\Omega$ is the angular velocity of the rotation. Since the 1970s, the Sagnac effect has found its crucial applications in navigation as the fundamental design principle of fiber optic gyroscopes (FOGs) [2,3]. In a FOG, when a single mode fiber is wound onto a circular coil with $N$ turns, the Sagnac effect is enhanced by $N$ times so that the travel-time difference is $2vNl/c^2$, where $v$ is the velocity of the rotating fiber, $l$ is the circumference of the circle, and $Nl$ is the total length of the fiber. The travel-time difference in a FOG can be expressed by the phase difference $\Delta\phi = 2\pi\Delta tc/\lambda$, where $\lambda$ is the free space wavelength of light. Today, FOGs have become highly sensitive detectors measuring rotational motion in navigation [4,5].

It is believed that the Sagnac effect exists only in circular motion. However, we have discovered that any moving path contributes to the total phase difference between two counterpropagating light beams in the loop. In a previous experiment using a fiber optic conveyor (FOC), we showed our preliminary result that a segment of linearly moving glass fiber contributes $\Delta\phi = 4\pi vL/c\lambda$ to the phase difference [6]. Here, we generalize our finding for the Sagnac effect with a more complete study and a series of new experiments. Our experiments include different types of motion and light paths with a glass fiber and an air-core fiber [7]. In this study, we found that in a light waveguide loop consisting of linearly and circularly moving segments, any segment of the loop contributes to the phase difference between two counterpropagating light beams in the loop. The contribution is proportional to a product of the moving velocity $\boldsymbol{v}$ and the projection of the segment length $\Delta\boldsymbol{l}$ on the moving direction, $\Delta\phi = 4\pi\boldsymbol{v}\cdot\Delta\boldsymbol{l}/c\lambda$. It is independent of the type of motion and the refractive index of waveguides. The total phase difference of the loop is $\Delta\phi = 4\pi\oint_l \boldsymbol{v}\cdot d\boldsymbol{l}/c\lambda$. This general conclusion includes the Sagnac effect for rotation as a special case. The finding also suggests a new fiber optic linear motion sensor having nanoscale sensitivity, which is much more sensitive than any existing linear motion detectors. This linear motion sensor may be applied to accelerometers in navigation and seismology.

Our experiments are different from the Fizeau-type experiment [8], "drag" experiment of a moving medium. In the Fizeau-type experiment, there is relative motion between the light path and the medium, water or glass. The result of the Fizeau-type experiment is dependent on the refractive index of the medium, and the drag coefficient is zero when the refractive index is 1. In our experiments, as in the FOG, there is no relative motion between the light path and the medium, glass fiber or air-core fiber.

The experimental system in our study consists of a fiber optic loop, a FOG [9], and a mechanical conveyor, as shown in Fig. 1(a). A single mode fiber loop with different configurations was added to the FOG, which was calibrated for the added length. The loop includes significant portions of fiber segments that move linearly. The function of the FOG in this experiment is to transmit and receive the counterpropagating light beams and to detect the phase differences between two light beams. Because the FOG is not rotating in the experiment, the detected phase differences are caused by the movement of the added fiber optic loop.

An air-core fiber was used to verify that the phase difference is independent of the refractive index and various types of motion. The new photonic band gap fiber [7] has a hollow air-guiding core for light with a wavelength of 1310 nm. The fiber was used to construct a two-wheel FOC shown in Fig. 1(a) and a three-wheel FOC, not shown. Shown in Fig. 1(b) are the detected phase differences caused by the fiber motion with both configurations. The length of the fiber loop is 4.1 m and the speeds of the motion are from 0.001 91 to 0.211 m/s. Each phase difference is an average of eight measurements. Thirteen phase differences range from 0.000 246 to 0.0276 rad. The maximum conveyor velocity error is 1.7%. The least-squares linear regression of the phase differences is $0.0323vL$, which closely matches with $\Delta\phi = 4\pi vL/c\lambda = 0.0320vL$ with $\lambda = 1310$ nm. It also matches the result using a glass fiber [6]. Therefore, it is confirmed that the contribution to the phase difference in a moving light waveguide is independent of the refractive index of the waveguide and independent of the loop configurations.

To study the relationship between the motion of the fiber and the fiber orientation, we conducted an experiment in which the fiber zigzags and has an angle $\theta$ with respect to the direction of fiber motion. Thus, for a fiber segment having an actual length of $\Delta l$, its effective length is $\Delta l\cos\theta$, which is a projection of the fiber onto the motion direction. As shown in Fig. 2, our experiment demonstrates that the effective length contributes the phase difference, not the actual length; therefore, the phase difference $\Delta\phi$ is not $4\pi v\Delta l/c\lambda$, but $4\pi v\Delta l\cos\theta/c\lambda = 4\pi \boldsymbol{v}\cdot\Delta\boldsymbol{l}/c\lambda$, which is the dot product between two vectors.

A further experiment was conducted to study the phase difference when different segments of the loop moved at different speeds. Figure 3(a) shows a fiber optic "parallelogram" where the top arm moves with the conveyor and the bottom arm is stationary. While moving, the two sidearms, being flexible, are kept the same shape so that the phase differences in these two sidearms cancel each other. There is no phase difference in the bottom stationary arm. Therefore, the detected phase difference is contributed solely by the motion of the top arm. The experiment was conducted using a glass fiber and an air-core fiber. The detected phase differences are shown in Fig. 3(b). For the glass fiber configuration, the length of the top arm is 1.455 m and there are 11 turns. The measured phase differences range from 0.000 482 to 0.121 rad for speeds from 0.000 917 to 0.233 m/s. In another configuration, the top arm is air-core fiber with a length $\Delta l$ of 5.23 m. The phase differences range from 0.000 301 to 0.0346 rad for speeds from 0.001 86 to 0.207 m/s. The least-squares linear regression of the phase differences is $0.0317v\Delta l$, which agrees with $\Delta\phi = 4\pi v\Delta l/c\lambda = 0.0320v\Delta l$ with $\lambda = 1310$ nm.

According to our experiments, we can draw a conclusion about the generalized Sagnac effect that in a moving fiber loop or waveguide, a segment $\Delta\boldsymbol{l}$ with a velocity $\boldsymbol{v}$ contributes $\Delta\phi = 4\pi\boldsymbol{v}\cdot\Delta\boldsymbol{l}/c\lambda$ to the total phase difference between two counterpropagating beams in the loop. The contribution $\Delta\phi$ is independent of the refractive index of the waveguide, and the motion of the segment can be either linear or circular. Thus, for the entire loop, the total phase difference between two counterpropagating beams in the loop is

$$\Delta\phi = 4\pi \oint_l \boldsymbol{v} \cdot d\boldsymbol{l}/c\lambda.$$

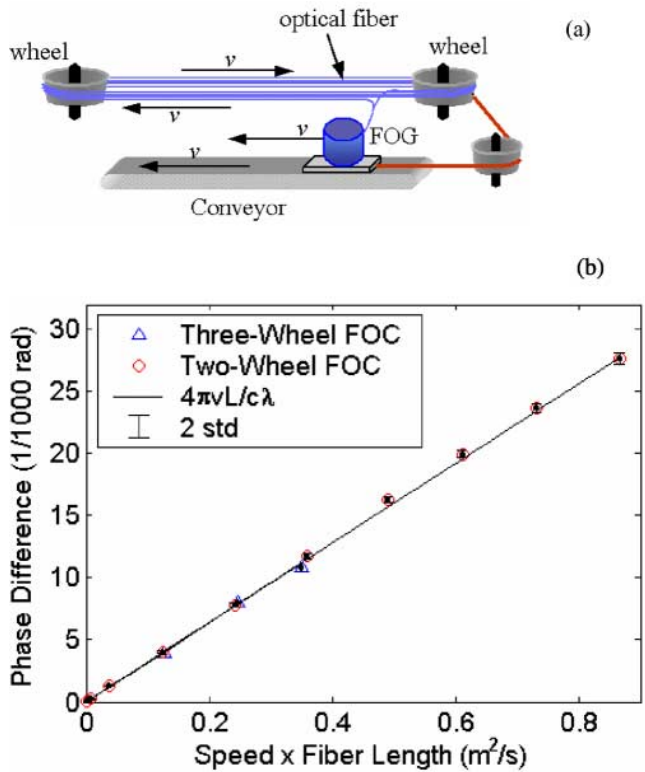

FIG. 1 Experiment for detecting the phase difference of two counterpropagating light beams in a moving fiber loop. (a) Experimental setup. The fiber loop is driven by the conveyor at a velocity $v$. The conveyor has a length of 1.5 m and can move from 0.001 to 0.25 m/s. The diameters of the wheels are 0.3 m. The FOG consists of a 1310-nm superluminescent light-emitting diode as the light source and a phase difference detector that has an output rate of 1162.6 mV per rad. (b) Phase differences of two counterpropagating beams in a moving air-core fiber.

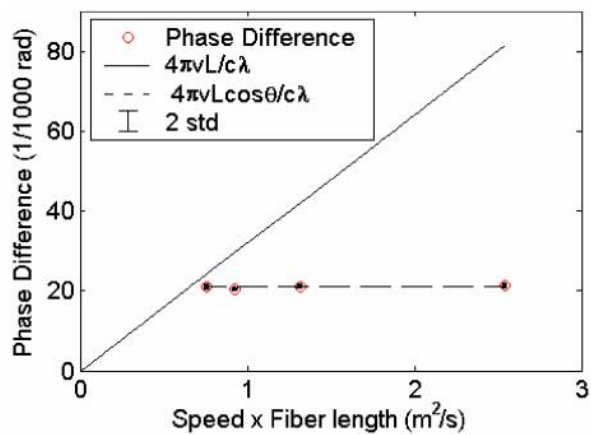

FIG. 2 Phase differences of moving fiber segments that have the same effective length and different actual lengths. The effective length is 7.20 m and four actual lengths are 8.31, 10.2, 14.4, and 27.8 m for $\theta = 30°$, $45°$, $60°$, and $75°$, respectively. The velocity is 0.091 m/s. Four phase differences range from 0.0206 to 0.0215 rad, with standard deviations from 0.000 280 to 0.000 743 rad. In this experiment, $4\pi v\Delta l\cos\theta/c\lambda = 0.032v\Delta l\cos\theta = 0.0211$ rad, which is represented by the dashed line.

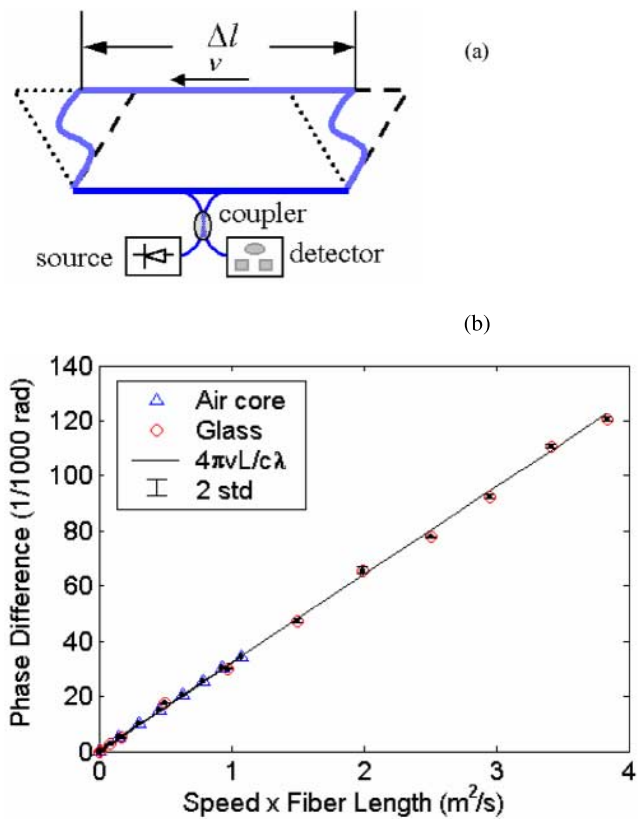

FIG. 3  Experiment for studying the phase difference when different segments of the loop move at different speeds. (a) Experimental setup. The light from a source is split into two beams that counterpropagate in the fiber which is wound onto a parallelogram. The bottom arm is fixed while the top arm is moving. The phase difference can be enhanced by multiple turns of the fiber on the parallelogram. The coupler, source, and detector are replaced by the FOG in the experiment. (b) Phase differences caused by the linear motion of the top arm of the fiber optic loop.

This general conclusion includes the Sagnac effect of rotation as a special case. In fact, the Sagnac result can be derived from this general equation by assuming a circular motion of the loop and using Stokes's theorem,

$$\Delta\phi = 4\pi \oint_l \boldsymbol{v} \cdot d\boldsymbol{l}/c\lambda = 4\pi \iint_A (\nabla \times \boldsymbol{v}) \cdot d\boldsymbol{A}/c\lambda$$
$$= 4\pi \iint_A 2\Omega \boldsymbol{k} \cdot d\boldsymbol{A}/c\lambda = 8\pi \Omega A/c\lambda.$$

This generalization provides a design principle for a new fiber optic linear motion sensor (FOLMS), which has a high sensitivity and a high stability. The basic structure of this sensor can be similar to that shown in Fig. 3(a). The linear motion of the top arm of the sensor is detected with a phase difference $\Delta\phi = 4\pi v N \Delta l/c\lambda$. Because two beams share the same optical path, the sensor is optically stable. Just as a FOG detects the rotational motion of an object, a FOLMS can detect the relative linear motion between two objects fixed on the top and bottom arms of the parallelogram. The optical technologies developed in recent decades for the FOG can be utilized for the FOLMS; therefore, the sensitivities and stability of the FOLMS are comparable to that of the FOG. The sensitivity of a FOG can be $10^{-7}$ rad of the phase difference [5]. With $N\Delta l = 500$ m and $\lambda = 10^{-6}$ m, a FOLMS can detect a linear velocity of

$$v = \Delta\phi c\lambda/4\pi N \Delta l = 4.8 \text{ nm/s}$$

which is a nanoscale velocity.

This nanoscale sensitivity linear motion sensor can detect the very small relative motion appearing in an accelerometer, which is important in navigation and seismology. The common design of an accelerometer is a spring-mass system. Improving the sensitivity of an accelerometer requires improving the sensitivity of detecting the linear movement of the mass relative to the base. Utilizing a FOLMS to detect the relative motion between the mass and the base will greatly increase the sensitivity of the accelerometer. It can be foreseen that an accelerometer using the FOLMS combined with the fiber optic gyroscope may be beneficial for navigation because both use the same technology and both are very stable optically.

We thank Dean Langley for useful discussions and help. We thank Robert Moeller of the Naval Research Laboratory for technical assistance and NRL for the loan of the FOG. The air-core fiber was purchased from Crystal Fiber, Denmark.